\documentclass[fleqn,usenatbib]{mnras}

\usepackage{newtxtext,newtxmath}

\usepackage[T1]{fontenc}

\DeclareRobustCommand{\VAN}[3]{#2}
\let\VANthebibliography\thebibliography
\def\thebibliography{\DeclareRobustCommand{\VAN}[3]{##3}\VANthebibliography}

\usepackage{graphicx}	
\usepackage{amsmath}	
\usepackage{aas_macros}
\usepackage{threeparttable}
\usepackage{xcolor}
\usepackage{twoopt}
\usepackage{orcidlink}
 
\newcommandtwoopt{\myfrac}[4][0pt][0pt]{\genfrac{}{}{}{}{\raisebox{#1}{$#3$}}{\raisebox{-#2}{$#4$}}}

\graphicspath{{imgs/}}

\newcommand\ergs{\ensuremath\mathrm{erg\,s^{-1}}}

\newcommand\ls{\ensuremath{\hbox{\rlap{\hbox{\lower4pt\hbox{$\sim$}}}\hbox{$<$}}}}
\newcommand\gs{\ensuremath{\hbox{\rlap{\hbox{\lower4pt\hbox{$\sim$}}}\hbox{$>$}}}}

\newcommand\ds{\ensuremath{D_\mathrm{S}}}
\newcommand\dls{\ensuremath{D_\mathrm{LS}}}
\newcommand\dlC{\ensuremath{D_\mathrm{L}^{\mathrm C}}}
\newcommand\dsC{\ensuremath{D_\mathrm{S}^{\mathrm C}}}
\newcommand\dlsC{\ensuremath{D_\mathrm{LS}^{\mathrm C}}}

\newcommand\zl{\ensuremath{z_{\rm L}}}

\newcommand\zs{\ensuremath{z_\mathrm{S}}}

\newcommand\mup{\ensuremath{\mu_{\rm p}}}
\newcommand\thE{\ensuremath{\theta_{\rm E}}}
\newcommand\tz{\ensuremath{\widetilde{z}}}
\newcommand\tD{\ensuremath{\widetilde{D}}}
\newcommand\tM{\ensuremath{\widetilde{\mathcal{M}}}}
\newcommand\tm{\ensuremath{\widetilde{m}}}

\newcommand\mR{\ensuremath{\mathcal{R}}}

\newcommand\Dt{\ensuremath{\Delta t}}

\newcommand\pgap{\ensuremath{p_{\rm gap}}}

\newcommand{\be}{\begin{equation}}
\newcommand{\ee}{\end{equation}}
\newcommand{\ba}{\begin{eqnarray}}
\newcommand{\ea}{\end{eqnarray}}

\defcitealias{lvc21}{LVK2021}

\title[Gravitational lensing interpretation of mass gap GWs]{On the gravitational lensing interpretation of three gravitational wave detections in the mass gap by LIGO and Virgo}

\author[M. Bianconi et al.]
{Matteo Bianconi\orcidlink{0000-0002-0427-5373},$\!^{1}\thanks{mbianconi@star.sr.bham.ac.uk}$
Graham P. Smith\orcidlink{0000-0003-4494-8277},$\!^{1}$
Matt Nicholl\orcidlink{0000-0002-2555-3192},$\!^{1,2}$
Dan Ryczanowski\orcidlink{0000-0002-4429-3429},$\!^1$
Johan Richard\orcidlink{0000-0001-5492-1049},$\!^3$
\newauthor
Mathilde Jauzac\orcidlink{0000-0003-1974-8732},$\!^{4,5,6,7}$
Richard Massey,$\!^5$
Andrew Robertson\orcidlink{0000-0002-0086-0524},$\!^8$
Keren Sharon\orcidlink{0000-0002-7559-0864},$\!^9$
Evan Ridley$^1$\\\\
$^1$School of Physics and Astronomy, University of Birmingham, Edgbaston, Birmingham, B15 2TT, UK\\
$^2$ Institute for Gravitational Wave Astronomy, University of Birmingham, Birmingham, B15 2TT, UK\\
$^3$ Univ Lyon, Univ Lyon1, Ens de Lyon, CNRS, Centre de Recherche Astrophysique de Lyon UMR5574, 69230, Saint-Genis-Laval, France\\
$^4$ Centre for Extragalactic Astronomy, Durham University, South Road, Durham, DH1 3LE, UK\\
$^5$ Institute for Computational Cosmology, Durham University, South Road, Durham, DH1 3LE, UK\\
$^6$ Astrophysics Research Centre, University of KwaZulu-Natal, Westville Campus, Durban 4041, South Africa\\
$^7$ School of Mathematics, Statistics \& Computer Science, University of KwaZulu-Natal, Westville Campus, Durban 4041, South Africa\\
$^8$ Jet Propulsion Laboratory, California Institute of Technology, 4800 Oak Grove Dr., Pasadena, CA 91109, USA\\
$^9$ Department of Astronomy, University of Michigan, 1085 S. University Ave, Ann Arbor, MI 48109, USA\\
}

\date{Accepted XXX. Received YYY; in original form ZZZ}

\pubyear{2020}

\begin{document}
\label{firstpage}
\pagerange{\pageref{firstpage}--\pageref{lastpage}}
\maketitle

\begin{abstract}
We search for gravitational wave (GW) events from LIGO-Virgo's third run that may have been affected by gravitational lensing. Gravitational lensing delays the arrival of GWs, and alters their amplitude -- thus biasing the inferred progenitor masses. This would provide a physically well-understood interpretation of GW detections in the ``mass gap'' between neutron stars and black holes, as gravitationally lensed binary neutron star (BNS) mergers. We selected three GW detections in LIGO-Virgo's third run for which the probability of at least one of the constituent compact objects being in the mass gap was reported as high with low latency -- i.e. candidate lensed BNS mergers. Our observations of powerful strong lensing clusters located adjacent to the peak of their sky localisation error maps reached a sensitivity $\rm AB\simeq25.5$ in the $z'$-band with the GMOS instruments on the Gemini telescopes, and detected no candidate lensed optical counterparts. We combine recent kilonova lightcurve models with recent predictions of the lensed BNS population and the properties of the objects that we followed up to show that realistic optical counterparts were detectable in our observations. Further detailed analysis of two of the candidates suggests that they are a plausible pair of images of the same low-mass binary black hole merger, lensed by a local galaxy or small group of galaxies. This further underlines that access to accurate mass information with low latency would improve the efficiency of candidate lensed BNS selection. 

\end{abstract}



\begin{figure*}
\begin{center}
    \includegraphics[width=0.5\linewidth, keepaspectratio, trim = 10cm 0 0 0 , clip]{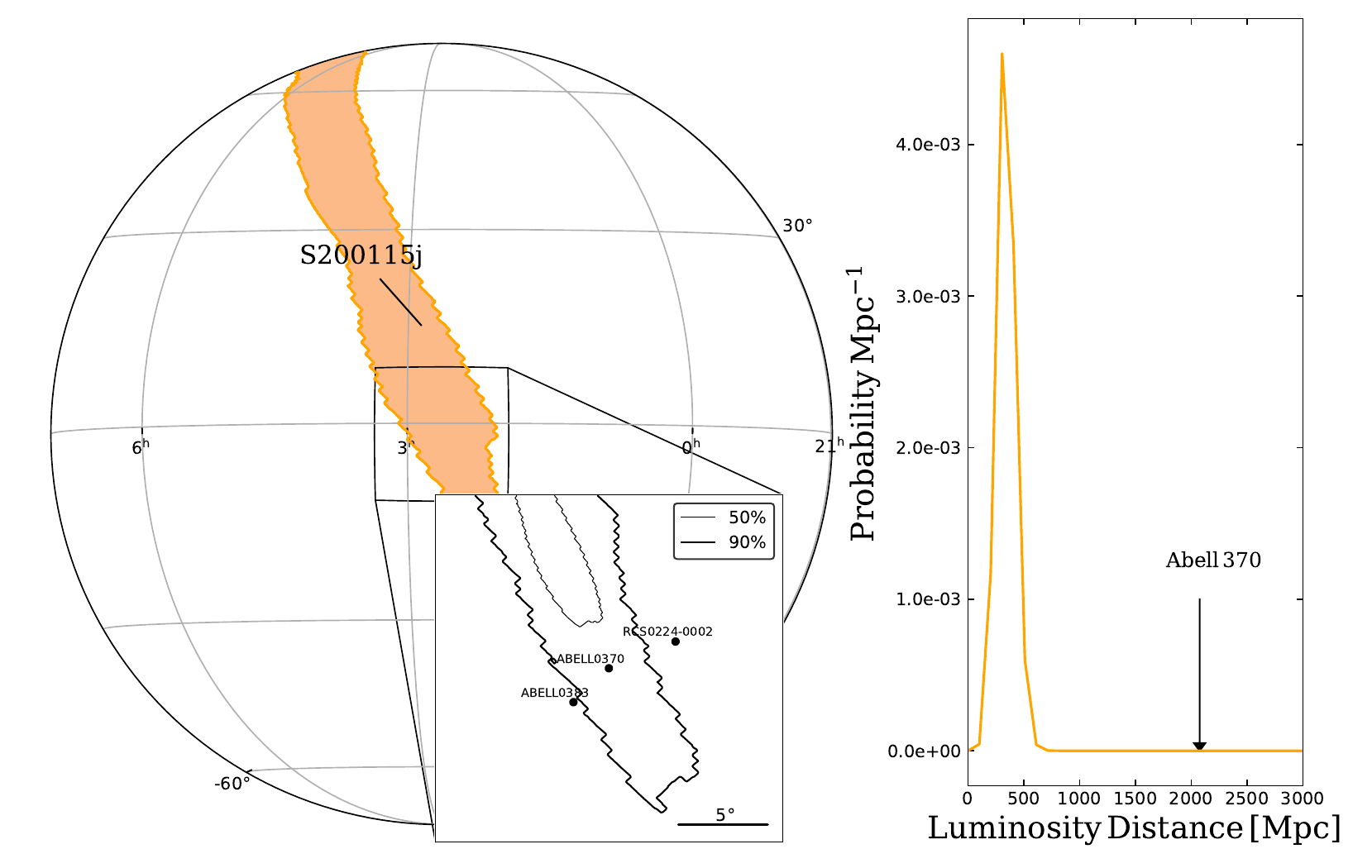}\includegraphics[width=0.5\linewidth, keepaspectratio, trim = 10cm 0 0 0 , clip]{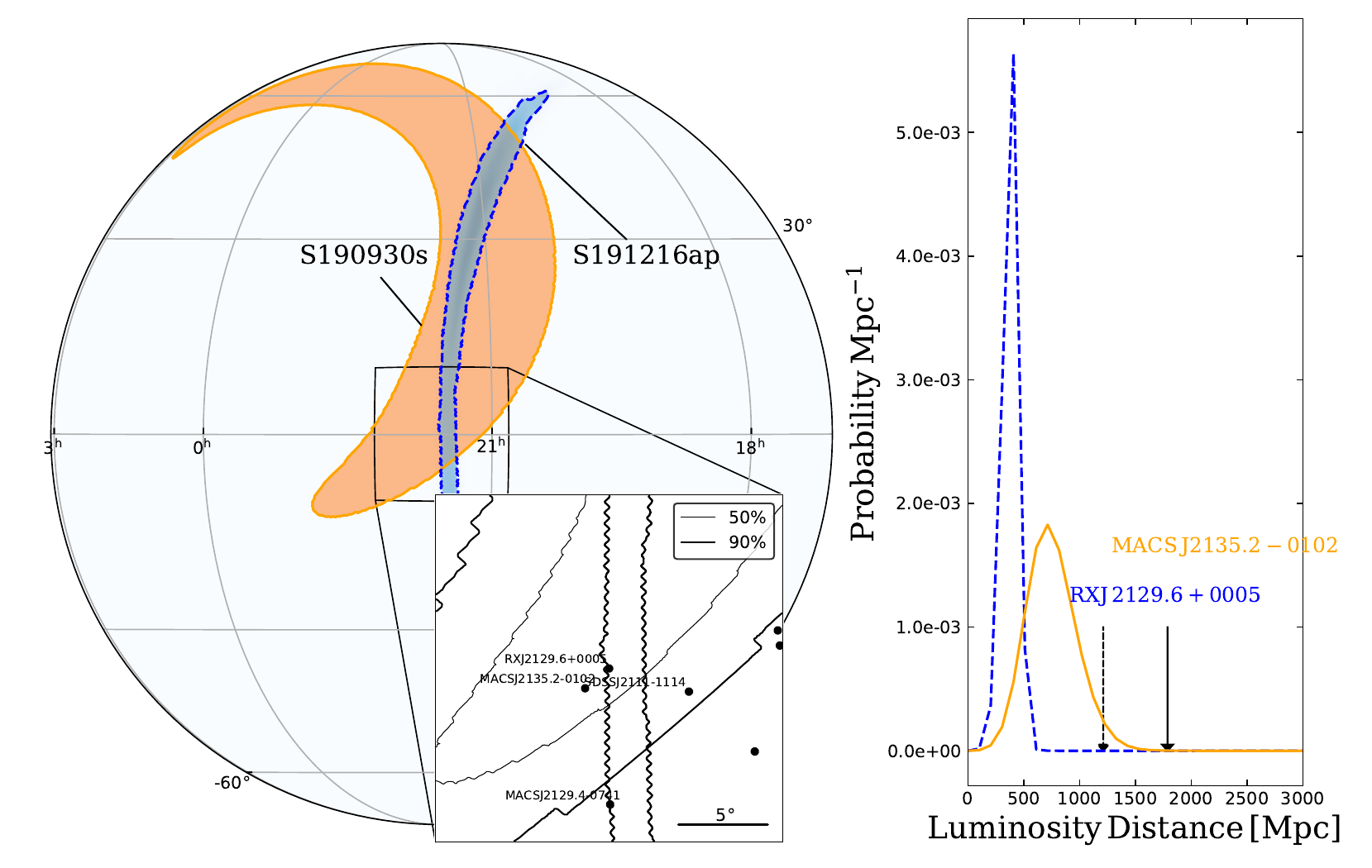}

 \end{center}
  \caption{Sky localisation in spherical projection of the candidate GW events considered in this work, together with their distance probability assuming no lensing intervention, and strong lensing cluster candidates from the \citet{smith18} list. The maps and distance posteriors are the ones available at the time of observations. {\sc Left}-- Spherical projection of the trigger S200115j skymap. The inset shows the location of the strong lensing cluster Abell 370, within the 90\% probability of the skymap. The lineplot highlights the probability distance of the event, with respect to the cluster redshift. {\sc Right}-- Same as left panel, for trigger S190930s and S191216ap, and the selected clusters MACS\,J2135.2$-$0102 and RX\,J2129.6$+$0005. See section~\ref{sec:target_selection} for additional details regarding the strong lensing cluster selection. } \label{fig:ligo_skymaps}
\end{figure*}
\section{Introduction}\label{sec:intro}

Gravitational lensing is contributing to unveiling otherwise inaccessible regions of the Universe. Intervening mass along the line-of-sight of conventional observations acts to magnify the observed radiation, allowing the detection of background objects whose direct electromagnetic (EM) radiation would be otherwise too faint because of their distance. This enabled the detection of increasingly remote galaxy populations (e.g. \citealt{kneib04, bouwens14}) and even individual stars at high redshifts (e.g. \citealt{kelly18, welch22}). Similarly, gravitational lensing can play a key role in investigating distant populations of GW sources, reaching beyond the limited sensitivity of current detectors, in particular in the low mass regime of binary neutron stars \citep[][and references therein]{smith21}. Therefore, it is highly relevant to investigate the scope for lensing as it currently provides the only window to low mass GW progenitors beyond the local Universe.

The census of confirmed GW events has steadily increased since the first detection \citep{abbott16} thanks to the continuous performance improvement of the network of instruments composing the LIGO, Virgo and KAGRA collaboration (LVK). The landscape of compact object mergers responsible for the GW emission is being continuously populated, in particular of progenitor masses between $1<m\,[{\rm M_\odot}]<100$. Interestingly, theoretical models of stellar evolution exclude the presence of compact objects in the mass interval between $2<m\,[{\rm M_\odot}]<5$, whose limits encompass the most massive neutron stars and the least massive black holes, respectively \citep{farr11,alsing18}. On the other hand, core-collapse supernova models can enforce this mass gap or produce a smooth remnant mass distribution by changing physical assumptions on the onset of the supernova explosion (e.g \citealt{Belczynski12, olejak20}). In their third run LVK detected candidate mass gap binary mergers hinting at different and varied formation channels responsible for the binary progenitors from the currently available in literature \citep{gupta20}. Alternatively, gravitational lensing offers a physically well-understood mechanism, whose intervention would cause the incorrect estimate of the GW progenitor parameters.

LVK classifies a GW source as a binary comprising of neutron stars (BNS), black holes (BBH), or neutron star-black hole (NSBH) system, according to the inferred mass retrieved from modelling of the detected waveform. The amplitude of the GW strain signal that the detectors measure, $A$, is affected by gravitational lensing magnification by intervening mass along the line-of-sight and the luminosity distance to the source, such that from the inverse square law: $A \propto \sqrt{\mu}/D$ \citep{wang96}. Here $\rm \mu$ is the magnification caused by gravitational lensing. The current LVK pipeline defaults to null gravitational lensing intervention, corresponding to magnification of $\rm \mu = 1 $. Therefore, if the GW source is actually strongly lensed (multiply imaged and thus $\rm \mu \gtrsim 2-10 $), LVK source distance posteriors will be biased low. Contextually, the inferred rest-frame mass $\tm$ of the compact objects responsible for the GW emission scales as $m(1+z)=\tm(1+\tz)$ where $m$ and $z$ are the true mass and redshift of the GW source, and $\tm$ and $\tz$ are the mass and redshift inferred with low latency assuming $\mu=1$. Hence, the masses inferred assuming $\rm \mu =1 $ require to be revised down if the source is lensed. This implies the released LVK event classification based on binary component mass could be biased (see also \citealt{smith21} for additional details). 

The LVK consortium has recently published a study on the impact of lensing on the GW detection rate and on searching for multiple images due to strong lensing within the events of the first half of O3, concluding against the occurrence of lensing \citep{abbott21}. This study relied on the analysis of the GW data stream by LVK. We argue for the need of
optical follow-up observations of candidate lensed GW sources in order to localise the source to a gravitational lens -- i.e. sub-arcsecond accuracy. This is currently not achievable with LVK data alone, despite the improvements in sky localisation achieved with three detectors \citep{abbott17c, abbott17a, abbott17b}. This motivates concentrating on lensed BNS  mergers because BNS are now confirmed as being associated with EM counterpart, i.e. kilonovae (KNe, e.g. \citealt{abbott17b}).  
In contrast, BBH mergers are expected to have no or very faint counterpart that is beyond the reach of today's telescopes (see also \citealt{graham20}), even when aided by gravitational lensing \citep[e.g.][and references therein]{smith19a}.

\begin{table*}
    \begin{center}
    \begin{tabular}{ c c c c c c c c c c}
         Name & Trigger & $p_{\rm gap}$ & $\tm_1$ $[M_{\odot}]$& $\tm_2$ $[M_{\odot}]$ & $\tM\,[M_{\odot}]$  & $\tD$ [Mpc] & \tz & SNR & 90\% Skymap area [${\rm deg}^2$] \\ 
         \hline
         \hline
        GW 190930\_133541 & S190930s & $>95\%$& $14.2^{+8.0}_{-4.0}$ & $6.9^{+2.4}_{-2.1}$ & $8.5^{+0.5}_{-0.4}$ & $770^{+320}_{-320}$ & $0.16^{+0.06}_{-0.06}$& $9.7^{+0.3}_{-0.5}$ & 1600 \\
        \hline
        GW 191216\_213338 & S191216ap & $>99\%$& $12.1^{+4.6}_{-2.3}$ & $7.7^{+1.6}_{-1.9}$ &   $8.33^{+0.22}_{-0.19}$ & $340^{+120}_{-130}$ & $0.07^{+0.02}_{-0.03}$ & $18.6^{+0.2}_{-0.2}$ & 490 \\
        \hline
        GW 200115\_042309 & S200115j &$>94\%$& $5.9^{+2.0}_{-2.5}$ & $1.44^{+0.85}_{-0.29}$& $2.43^{+0.05}_{-0.07}$ &$290^{+150}_{-100}$& $0.06^{+0.03}_{-0.02}$ & $11.3^{+0.3}_{-0.5}$ & 370 \\
        \hline
    \end{tabular}
        \end{center}
    \caption{Summary of salient posteriors and their 90$\%$ confidence interval of the confirmed GW events considered in this study, from LVK data release assuming no lensing. From left to right, name of the confirmed GW event, trigger ID, mass gap probability released by LVK with low latency, individual masses and chirp mass of the binary, luminosity distance, redshift, network match-filtered SNR, and credible area of the sky localisation  from \citet{abbott21.1} and  \citetalias{lvc21}.}\label{table:gw_event}
\end{table*}

If on the one hand lensing masks intrinsic properties of the GW binary, on the other hand it can be used to search further along the recesses of the Universe for the EM counterpart of the GW event. Simple lensing arguments show the impact of magnification on the expected observed magnitude of the electromagnetic counterpart of a BNS. For example, consider the detection of a GW170817/AT2017gfo-like counterpart to a BNS merger located at $z\simeq1$ that is lensed by a massive galaxy or cluster in the foreground. Within four observer-frame days from the GW event, such an object would have absolute magnitude of $M\lesssim-14$ \citep{arcavi18}. If the EM counterpart to a gravitationally lensed BNS merger located at $z\simeq1$ has a similar luminosity, then its apparent magnitude is given by: $m\simeq-14+5\log(D/10{\rm pc})-2.5\log(\mu)$ where $D$ is the intrinsic luminosity distance to the lensed GW, $\mu=(D/\tD)^2$ is the gravitational magnification suffered by the source, and $\tD$ is the
luminosity distance initially assigned to the source by LVK
assuming $\mu=1$. Interestingly, $m\simeq-14+5\log(\tD/10\,{\rm pc})$, because the inverse square law cancels the gravitational magnification and $k$-corrections appear to be modest \citep{smith21}. Assuming that the GW is initially placed at $\tD=600\,{\rm Mpc}$, this implies an apparent magnitude of $m\simeq25$. Once a typical strong lensing magnification for lensed BNS mergers of $\mu = 100$ \citep{smith21} is taken into consideration, the effective magnitude that can be accessed reaches $m + 2.5\log(\it{\mu})$, equalling $m\approx 30$. Therefore, lensing is crucial not only to boost the EM counterpart signal, but also necessary to access BNS mergers at redshift $z\gtrsim1$.

This paper is organised as follows.  
In Section~\ref{sec:strategy} we motivate our observational strategy and summarise the GW events from LVK's third run that we have targeted in our follow-up campaign. In Section~\ref{sec:gmos} we describe the data reduction and analysis. In Section~\ref{sec:models} we introduce the assumptions and models used for assessing the impact of lensing on the detectability of BNS mergers and their EM counterparts, and present the models of KN lightcurves that we use to validate our follow-up observations. In Section~\ref{sec:results} we discuss the physical interpretation of our observations and the candidates that we observed. We close by summarising our results in Section~\ref{sec:conclusions}, and by discussing the impact of posterior parameters on the observing strategy of future campaigns. We assume cosmology values presented in \citet{planck16}, with $h=0.678$, $H_0 = 100\, h\, \rm km\,s^{-1}\,Mpc^{-1}$, $\Omega_{\rm M} = 0.309$ and $\Omega_\Lambda = 0.691$. All celestial coordinates are stated at the J2000 epoch, and all magnitudes are stated in the AB system.

\section{Campaign rationale}\label{sec:strategy}
Our efforts are concentrated on identifying GW sources that may be affected by strong lensing. Therefore, we devised a selection criterium aiming at events that LVK initially classifies as having at least one progenitor component in the mass gap with a powerful strong lensing cluster located within its
sky localisation. Mass gap progenitors are compact object with masses between $2.5< m [{\rm M_\odot}]<5$, which encompass the mass boundaries of the most massive neutron star (NS) and the lightest black hole (BH) known \citepalias{lvc21}. Hence, any object within this mass range cannot be accounted for by current models of stellar evolution. A natural solution to the detection of these object is invoking the impact of strong lensing, which would bias the recovered posterior masses. Recent studies from numerical simulations show that
galaxies ($\rm M_{200} < 10^{13}\,M_{\odot}$) and clusters ($\rm M_{200} > 10^{13}\,M_{\odot}$) contribute roughly equally to the optical depth to strong lensing \citep{robertson19}. We focus our effort on known strong lensing clusters primarily because of the capabilities of current observational infrastructures. A search of all galaxy- and cluster-scale lenses would require $\ge 60$ hours of observing with Subaru/HSC or CTIO/DECam to cover the full median sky localisation error region of $\gtrsim 100\,\rm degree^2$ to the necessary depth. This is stretching beyond the limit of observing capabilities at the current larger observational facilities, before the beginning of the Vera Rubin Observatory's Legacy Survey of Space and Time. Hence, we focus on the most promising line of sight, i.e. cores of strong lensing clusters, which typically cover few square arcminutes and account for $\simeq5$ per cent of the optical depth to strong lensing per GW sky localisation \citep{robertson19}. We combine this with \citeauthor{smith21}'s (\citeyear{smith21}) detailed predictions of the rate of lensed BNS detections in O3 ($0.02\,\rm yr^{-1}$ for their baseline model) to estimate the probability of successfully confirming a candidate lensed BNS with detection of its EM counterpart with our strategy of $\approx0.1\%$. We refer the interested reader to and \citet{robertson19} and \citet{smith21} for full details of the underlying calculations.

\begin{figure*}
\begin{center}
    \includegraphics[width=0.33\linewidth, keepaspectratio]{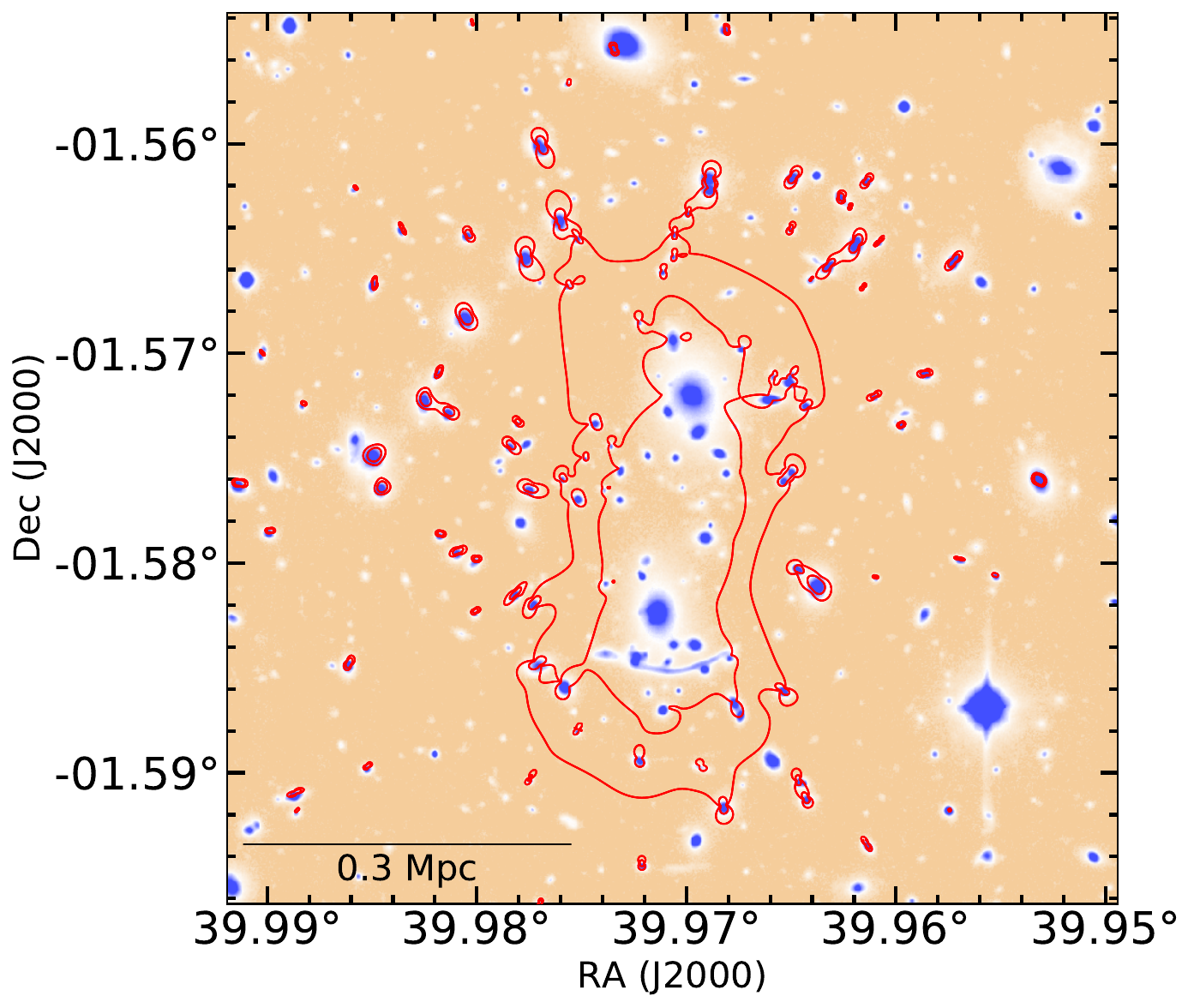} \includegraphics[width=0.33\linewidth, keepaspectratio]{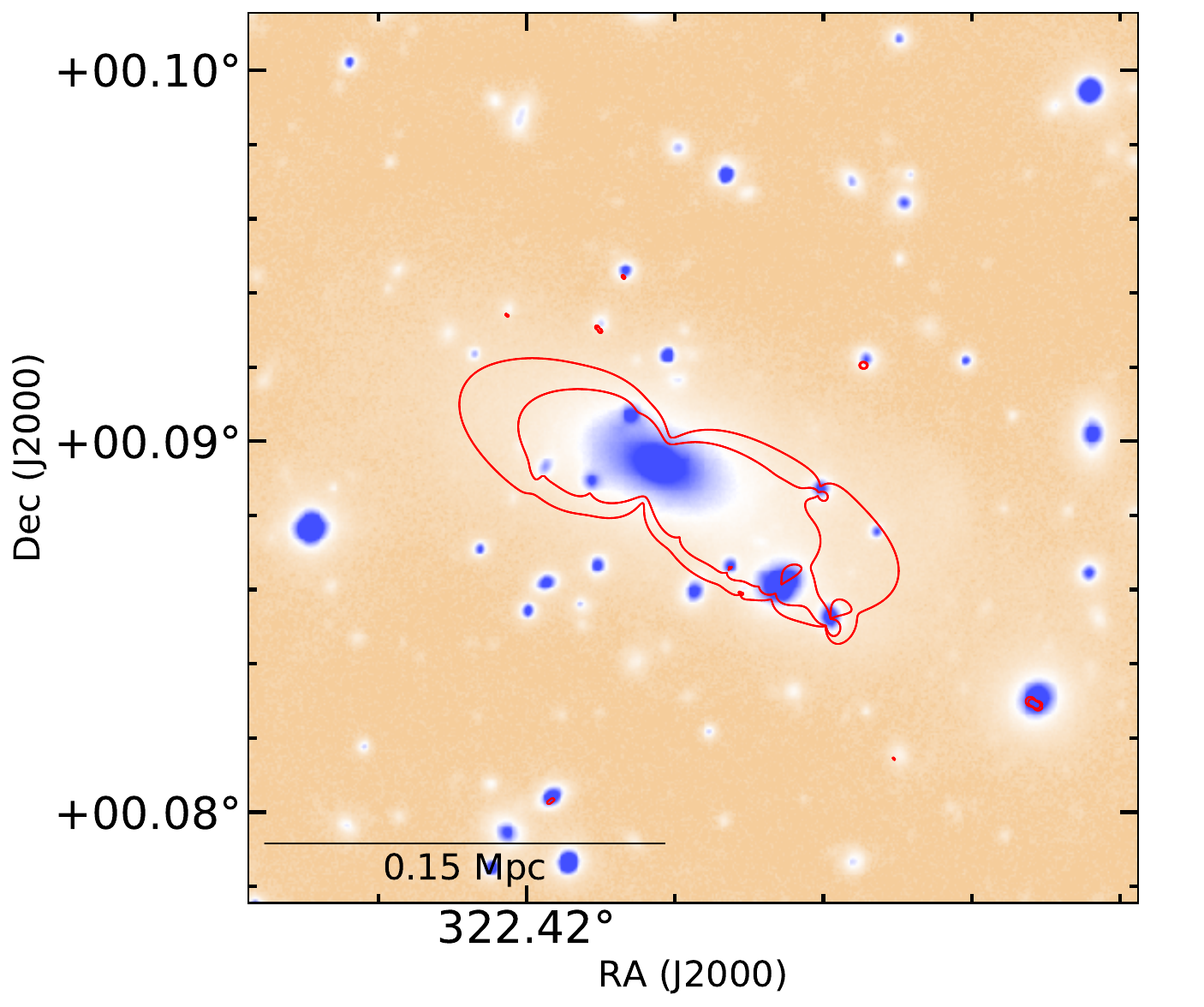}\includegraphics[width=0.33\linewidth, keepaspectratio]{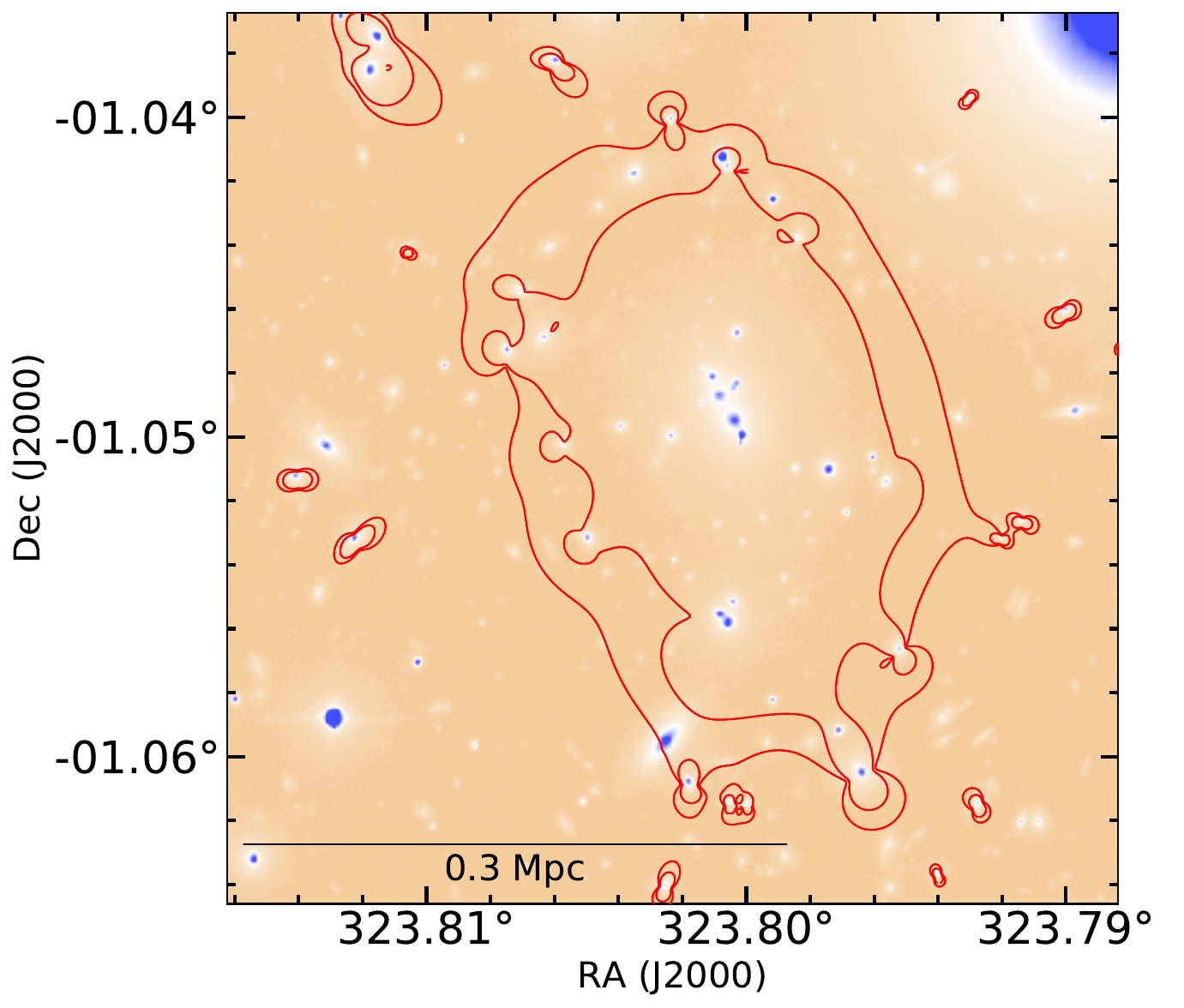}
   
 \end{center}
  \caption{Central cutouts of the Gemini GMOS $z'$-band images of Abell 370 (left panel), RX\,J2129.6$+$0005 (middle panel) and MACS\,2135.2$-$0102 (right panel). North is up and East is to the left. Critical curves are plotted as red lines, and mark the location of infinite magnification for source redshifts of $z=1$ and $z=2$.} \label{fig:gmos_images}
\end{figure*}

\subsection{Observing strategy and target selection}\label{sec:target_selection}

Our intent is to obtain new observations of strong lensing cluster cores soon after the detection of a GW event, to search for EM transients by comparing them to archival datasest and to communicate the finding within the transient community to extend the follow-up efforts. 
We require rapid observations because KNe fade quickly, therefore observations as soon as possible after the LVK detection are essential. Our previous attempts at detecting lensed EM counterparts to GW events \citep{smith19a}, and extensive follow-up campaign of the KN AT2017gfo associated to GW170817 \citep{villar17} informed our choice of instrument, waveband and observing time. As a result, we chose redder optical wavebands, in which the EM emission from the KN is more luminous and longer-lasting \citep{arcavi18}. In particular, we converged on 1-hour integration time and $z$-band filter (central wavelength $\approx 900$ nm) which allows to reach magnitude limits $m\approx25$ depth using a 8-meter class ground-based telescope (see discussion in Section \ref{sec:intro}).

Our list of galaxy clusters acting as gravitational telescopes comprises the sample of 130 strong lensing clusters discussed by \citet{smith18}. Thanks to numerous \emph{HST} and 8-meter class ground-based telescope observing campaigns that have targeted X-ray luminous galaxy clusters (e.g. \citealt{kneib96, smith05,richard10, jauzac16, lagattuta17, mahler18}), these cluster lenses are all spectroscopically confirmed, with a well constrained model of the cluster mass distribution. The choice of the most promising cluster to observe follows the identification of the closest object to the peak of the 2d probability distribution of each GW sky localization. Additional cluster characteristics, in particular Einstein radius size, are considered in the case of multiple clusters being available to observe.

\subsection{Candidate events with possible EM counterpart}\label{sec:massgap_events}

Hereafter we summarise the mass gap events detected by LVK's third run, and our strategy to select the most promising ones, and to observe the cluster targets within the localisation maps of the GW events. These events, which are listed in Table~\ref{table:gw_event} have been selected for having a high mass gap probability (>94\%) and a low false alarm rate at the time of announcement. 

LVK announced the detection of six GWs with $\pgap>90\%$ during the period overlapping with our access to Gemini, from Spring 2019 to Spring 2020. We summarise these six briefly here, in chronological order. S190426c was assigned high probability of originating from a BNS or a mass gap event (GCN24250, \citealt{smith19gcn}), but its sky localisation shared no overlap with our galaxy cluster list. S190924h was classified as mass gap event which we didn't consider due to the sub-optimal position on the sky, resulting in estimated high airmass for our telescope pointings. This was coupled with delays in the release of the offline sky localisation from LVK, which resulted in us prioritizing the subsequent event S190930s instead. S190930s, together with S191216ap and S200115j, were well suited for our EM counterpart search, and are the targets of our observations. Lastly S200316bj was communicated after our share of observing time at Gemini had been used. The target decision tree is detailed in the following subsections for each GW event we considered in this study. A visual representation of each GW sky localisation at the time of observations, together with the most promising strong lensing cluster candidates is presented in Figure~\ref{fig:ligo_skymaps}. The main properties of clusters that we have selected is presented in Table~\ref{tab:clusters}.

\subsubsection{GW190930\_133541}
The event (announced as S190930s) was detected by both LIGO L1 and H1 on September 30, 2019 at 13:35:41 UTC with an initial false alarm rate of 1/10.52 years (GCN Circular 25871). The \texttt{BAYESTAR} pipeline \citep{singer16} assigned 50\% probability of the sky localization spanning $706\, \rm{deg^2}$, mass gap probability >95\% and distance 752$\pm$224 Mpc. We located 4 strong lensing clusters from our list within the 90\% skymap, including MACS\,J2135.2$-$0102 (redshift z = 0.33). This cluster lays on the contour enclosing p = 0.177 of the localization probability density and is the closest cluster to the peak of the skymap, corresponding to a region subtending 196.92 $\rm{deg}^2$, and has an Einstein radius $\theta_E$= 38 arcsec, the largest among the four clusters. Subsequently, the 50\% skymap and distance were updated to $536 \, \rm{deg^2}$ and 709$\pm$191 Mpc by the \texttt{LALINFERENCE} pipeline \citep{veitch15}, in which only two strong lensing clusters were contained. MACS\,J2135.2$-$0102 lays on the contour enclosing p=0.825 per cent corresponding to a region subtending Area=1556.31 $\rm{deg}^{2}$. This map was released prior to our observations, in which we have triggered Gemini-North observations totalling 1-hour in the $z'$-band (program ID:GN-2019B-Q-205). The GW event has been subsequently confirmed by several offline pipelines and included in the LIGO/Virgo source catalogue \citep{gwtc2}. The final skymap (50\% covering  $569\, \rm{deg^2}$, distance 770$\pm$320 Mpc, source masses $m_1= 14.2 \rm M_{\odot}$ $m_2 = 6.9 \rm M_{\odot}$) includes MACS\,J2135.2$-$0102 on the contour enclosing p=0.680 per cent corresponding to a region subtending Area=924 $\rm{deg}^{2}$. The false alarm rate has been updated to of $\approx$ 1/30 years. Interestingly, the only other cluster consistently selected by our search scheme was RX\,J2129.6$+$0005,  across the refinement process of the sky localisation maps (see following section).

\begin{table*}
  \caption{Summary of salient properties of the clusters Abell~370, MACS\,J2135.2$-$0102 and RX\,J2129.6$+$0005 located within the skymaps of the GW events listed in Table~\ref{table:gw_event}}
  \label{tab:clusters}
  \begin{threeparttable}
    \begin{tabular}{p{30mm}p{18mm}p{25mm}p{18mm}}
      \hline 
       & Abell~370 & MACS\,J2135.2$-$0102 & RX\,J2129.6$+$0005 \cr
      \hline 
      Cluster redshift & $\rm 0.375$ & $\rm 0.33$ & $\rm 0.235$\cr
	  \noalign{\smallskip}
	  Right ascension & $\rm 02^h 39^m 52.9^s$ & $\rm 21^h 35^m 15.192^s$ & $\rm 21^h 29^m 39.6^s$ \cr
	  \noalign{\smallskip}
	  Declination & $-01^{\circ} 34' 36.5''$ & $-01^{\circ} 03' 01.70''$ & $00^{\circ} 05' 21.2''$\cr
	  \noalign{\smallskip}
	  \raggedright $M_{500}$ [$10^{14}\, M_{\rm \odot}$] & \raggedright $10.6 $\tnote{a}  & \raggedright $7.6$\tnote{b} & $\rm 3.5$\tnote{c}\cr
	  \noalign{\smallskip}
	  \raggedright $L_X^{\rm bol}$ [$10^{44}\ergs$] & \raggedright $11.1 $\tnote{a}  & 4.1 \raggedright \tnote{d} & $\rm 21.1$\tnote{c}\cr
	  \noalign{\smallskip}
	  \raggedright Einstein radius [arcsec] & 39 ($z_s$ = 2)\tnote{e} & 38 ($z_s$ = 2)\tnote{e}  & 18 ($z_s$ = 2)\tnote{e}\cr

	  \noalign{\smallskip}
	  \hline
    \end{tabular}
    \begin{tablenotes}
      \item [a] \citet{morandi07}
      \item [b] \citet{giacintucci17}
      \item [c] \citet{Okabe2016}
      \item [d] \citet{hlarrondo12}
      \item [e] \citet{richard10}
    \end{tablenotes}
  \end{threeparttable}
\end{table*}

\subsubsection{GW191216\_213338}

The event (announced as S191216ap) was detected on December 16 2019 at 21:33:38 UTC by the online pipeline (\texttt{BAYESTAR}) analysing the data stream from LIGO H1 and Virgo. The event presented a false alarm rate of $\approx$ 1 in $10^{15}$ years, a $50\%$ skymap encompassing 85 $\rm{deg}^2$, mass gap probability >99\% and distance $324\pm78$ Mpc. Our finder routine identified two strong lensing clusters within the $90\%$ probability of the event peak, of which RX\,J2129.6$+$0005 (redshift $z = 0.23$, Einstein radius $\theta_E$ = 17") was closest to the peak and located within p=0.717 corresponding to an  area = $158.29\,{\rm deg}^2$. An updated skymap from the \texttt{LALinference} pipeline showed the $50\%$ probability to 68 ${\rm deg}^2$, false alarm rate of  $\approx$ 3 in $10^{15}$ years, and a distance of $376\pm70$ Mpc. The cluster RX\,J2129.6$+$0005 was the only cluster left from our search at p=0.876 corresponding to an area = $226.05 \,{\rm deg}^2$. Furthermore, at $\tD\simeq300\,\rm Mpc$, this detection is close to the peak of the predicted population of lensed BNS mergers in LVK's third run \citep{smith21}.  Following the GW trigger announcement, two GCN circulars confirmed the detection of a contextual neutrino by IceCube (GCN 26460) and a sub-threshold gamma-ray source by HAWC (GCN 26472) from a sky position offset from RX\,J2129.6$+$0005 \citep{icecube19, hawk19}. Therefore, we applied for additional Director Discretionary observing time at UKIRT to image the location of the HAWC detection, as detailed in Section \ref{sec:ukirt}.

\subsubsection{GW200115\_042309}

The event (announced as S200115j) was detected on January 15, 2020 at 04:23:09.742 UTC with false alarm rate of $\approx$ 1 in $10^3$ years and a mass gap probability > 94\%. The initial skymap produced by the online pipeline \texttt{BAYESTAR} was updated three times within a few hours of the initial event release. As suggested in the LIGO/Virgo GCN 26759, the third skymap release was the preferred choice presenting a 50\% sky probability of 186 ${\rm deg^2}$. The mass gap probability was updated to > 99\% and distance $331\pm97 \, {\rm Mpc}$. Within this skymap, we identified three strong lensing clusters including Abell 370 (z=0.38, Einstein radius: 45"). The cluster is the closest to the sky localisation peak and located at p=0.163 corresponding to an  area = 31.53 ${\rm deg}^2$. Thanks to the large Einstein radius of the cluster, together with the plethora of ancillary archival data which includes deep multiband \emph{HST} imaging, serving as reference image for our science case, we triggered a 1-hour observation with Gemini South in the $z'$-band (program ID: GS-2020A-Q-136). A subsequent skymap update from the ${\rm LALinference}$ resulted in a 50\% sky probability of 153 ${\rm deg^2}$, distance $340\pm79 \, \rm Mpc$ and false alarm rate $\approx$ 1 in 1500 years, in which Abell 370 was the only cluster identified as located at p=0.595, corresponding to an area = 222.76 ${\rm deg}^2$.

\section{Data collection and analysis}\label{sec:gmos}

Our choice to use the Gemini telescopes is motivated primarily by the 5.5 sq. arcmin field of view of the GMOS instrument, which is large enough to cover to the typical angular extent of the strong lensing region of the clusters in our sample, together with the capability of accessing both northern and southern sky. In addition, we made use of Director Discretionary Time at the UKIRT telescope to access the WFCAM instrument, whose field of view is well-matched to the size of the ICECUBE localisation uncertainty of the neutrino detection discussed above.

\subsection{Gemini observations}
Our target of opportunity (ToO) observing programmes at the Gemini Observatory covered the duration of  LVK's third run, from April 2019 to May 2020 (Program IDs GN-2019B-Q-205, GN-2020A-Q-139 and GS-2020A-Q-136; PI: Bianconi). Each of these programs allowed for a single 1-hour visit with the GMOS instrument on the Gemini-North and South telescopes. Table~\ref{tab:obs} summarises the conditions and integration time of the individual observations. We have applied the same data reduction pipeline to each cluster observation, as described hereafter.
Individual GMOS exposures were de-biased, dark-subtracted, flat-fielded, and de-fringed using Gemini {\sc dragons} Python package\footnote{https://dragons.readthedocs.io/projects/gmosimg-drtutorial/en/stable/index.html}, to produce a single science frame per exposure comprising the mosaiced individual chips. The individual exposure frames are stacked to produce a single frame per visit, after masking bad pixels. The full width at half maximum of point sources in the reduced frames is consistently sub-arcsecond in all our observations (see Table~\ref{tab:obs}). The central cutouts of the GMOS images are presented in Figure~\ref{fig:gmos_images}.

\subsection{UKIRT observations}\label{sec:ukirt}
The GW trigger S191216ap was followed by a neutrino detection obtained by IceCube, whose sky localisation is consistent with that of the GW event \citep{icecube19}. Subsequently, a  sub-threshold gamma ray detection was announced by the HAWC collaboration, with a sky localisation consistent with both LIGO and IceCube \citep[][]{hawk19}, but offset with respect to the position of the cluster RXJ 2129.6+0005, which had been targeted by our Gemini observations. Therefore, we identified a circle of radius 0.3 degrees encompassing 68\% probability of the HAWC detection, and obtained Director Discretionary time observations (Program ID: U/19b/D05) with the WFCAM instrument on UKIRT through the $z$-band covering 0.75 sq. degrees. We performed an 1-hour long observation centred at RA 21:32:00 Dec +05:13:48 on 2019-12-20, characterised by airmass $\approx$1.5, FWHM$\approx$ 1.4 arcsec and a $5\sigma$ magnitude limit $m_z\approx21.5$. A second 1-hour visit was completed on the following day, but yielded lower magnitude limits with respect to the previous night, hence rendering the search for a fading transient difficult. From the comparison between the Epoch 1 and Pan-STARRS1 archival data, we identified two candidate transients not associated with any known or candidate gravitational lens, one of which is located $\approx 7$ arcsec from an edge-on galaxy at RA: 21:32:45.97 Dec:+5:19:57.0 (GCN 26605, \citealt{smith19gcnb}). Further analysis highlighted the probable bogus nature of these candidates, due to detector-related artefacts.

\subsection{Comparison with archival \emph{HST} observations}

The search for candidate lensed EM counterparts was performed by means of visual identification of new sources in proximity to the strong lensing regions of the clusters considered here. Archival \emph{HST} WFC3-IR images through filters F105W and F110W were used as reference for the search of new transients in the GMOS frames. This search was performed by several participant of the collaboration promptly after the collection of the new data. The depth reached by the archival \emph{HST} imaging is listed hereafter for each cluster at the $5\sigma$ level. Stacked \emph{HST} images of Abell~370, MACS~J2135.2$\_$0102 and RX~J2129.6$+$0005 reach $m\approx 29$ (Proposal ID: 14038, PI: Lotz), $m\approx 25.5$ (Proposal ID: 12166, PI: Ebeling, filter F110W), $m\approx 27$ (Proposal ID: 12457, PI: Postman), respectively.
\begin{table*}
\caption{Follow-up observations of strong lensing clusters within sky
  localizations of the GW 190930\_133541, 191216\_213338 and 200115\_042309.}
\label{tab:obs}
\centering
\begin{threeparttable}
\begin{tabular}{cccccc}
      \hline 
      Visit & Start of observation (UTC) & Airmass\tnote{a}    & Integration & Seeing\tnote{b}   &  Sensitivity\tnote{c} \cr 
            &                            &       & time (ks)   & (arcsec) & \cr

      \hline 
      \multispan6{\sc \hfil GMOS-N Observations of MACS\,J2135.2$-$0102\hfil}\cr
      \hline
      1 & October 07, 2019, $07{:}31{:}44$ & $1.12$ & $2.7$ & $0.68$ & $25.5$ \cr
      \hline
      \multispan6{\sc \hfil GMOS-N Observations of RX\,J2129.6$+$0005\hfil}\cr
      \hline
      1 & December 19, 2019, $04{:}47{:}55$ & $1.61$ & $2.7$ & $0.78$ & $25.5$ \cr
      \hline
      \multispan6{\sc \hfil GMOS-S Observations of Abell~370\hfil}\cr
      \hline
      1 & January 16, 2020, $00{:}58{:}52$ & $1.27$ & $2.7$ & $0.76$ & $25.4$ \cr 

      \hline
    \end{tabular}
    \begin{tablenotes}
    \item [a] The airmass at the mid-point of the observation.
    \item [b] Mean full width at half maximum of point sources in the
      reduced data.
    \item [c] $5\sigma$ point source sensitivity within a photometric
      aperture of diameter $2\,{\rm arcsec}$, estimated from the
      magnitude at which the photometric uncertainty is
      $0.2\,{\rm magnitudes}$.
    \end{tablenotes}
\end{threeparttable}
\end{table*}

\subsection{Photometric calibration}

The angular extent and depth of the GMOS frames meant that there was no overlap between unsaturated bright stars in our field-of-view and measured in all-sky surveys. Therefore we benchmarked the photometric calibration by measuring the ($g'-z'$) colours using archival SDSS catalogues and comparing it with model colours for massive early-type galaxies at the clusters' redshift. These model colours were computed using the EzGal code \footnote{www.baryons.org/ezgal}, and considering a single stellar population that formed at high redshift and evolved passively to the relevant cluster redshifts based on the \citep{bruzual03} populations.  Formation redshift and the metallicity do not impact significantly the predicted colours. We obtained a consistent 5 $\sigma$ sensitivity level across each cluster observations reaching $m\approx25.5$ in the $z'$-band.

\section{Lensing and EM radiation models}\label{sec:models}
The following sections introduce the main aspects of the theoretical framework used to predict and interpret the signatures of EM lensed counterparts in our observations. We summarise the methods used to compute BNS rates as a function of mass and redshift, the delay time expected for a GW event as a function of lens properties, and state-of-the-art model lightcurves for KNe. We refer the reader to \citet{smith21} and \citet{nicholl21} for a complete description of the models considered here.

\subsection{GW progenitor model}\label{sec:gw_model}

The baseline model assumes a rate of BNS mergers as follows:
\begin{equation}
  \mR(z,m)=\mR_0\,\,g(z)\,\,f(m)
  \label{eqn:mRmodel}
\end{equation}
where $\mR_0$ is the comoving merger rate density of BNS mergers in the local universe (\citetalias{lvc21}).  The functions $g(z)$ and $f(m)$ regulate the redshift evolution of GW sources, following the typical cosmic star formation rate density \citep{madau14}, and the BNS mass function respectively. The latter takes the shape of a top-hat function within the mass interval $1<m\, [{\rm M_\odot}] <2.5$, following \citetalias{lvc21}. 

The number of lensed GWs per year arriving at Earth can be obtained by integrating the rate of BNS mergers $\mR(z,m)$ along redshift. In doing so, the differential source-plane optical depth is included, $\tau^{\rm S}_\mu$, which describes the fraction of the source plane that is magnified via lensing. A lower limit on redshift $z_0$ is imposed and corresponds to the distance $D_0$ for a magnification yielding multiple images, i.e. $\mu=(D_0/\tD)^2=2$. Finally, the sensitivity of the GW detectors is taken into account \citep{martynov16,chen21} in order to capture those lensed GWs that arrive at Earth that are detectable by current instruments.

The lensing model also permits predictions of the time delay between the arrival of multiply lensed GWs. Qualitatively, the number of images a source generates depends on the structure of the gravitational potential of the lens, and the alignment between source and lens caustics, i.e. theoretical surfaces corresponding to infinite magnification. \citet{smith21} showed that the pseudo catastrophe of the singular isothermal lens and the fold catastrophe bracket the range of time delays measured to date for lensed quasars, and thus form a solid basis for predicting time delays for other lensed transients. We therefore adopt the following convenient scaling relations for lenses with an isothermal slope:
\begin{equation}
    \frac{\Dt_{\rm SIS}}{92\,\rm days}=\left[\frac{\thE}{1''}\right]^2\,\bigg[\frac{\mup}{4}\bigg]^{-1}\,\left[\frac{\mathcal{D}}{3.3\,\rm Gpc}\right]
    \label{eqn:tdouble}
\end{equation}
\begin{equation}
    \frac{\Dt_{\rm fold}}{3.9\,\rm days}=\left[\myfrac[0pt][1pt]{\thE}{1''}\right]^2\left[\myfrac[0pt][1pt]{\mup}{4}\right]^{-3}\left[\myfrac[0pt][1pt]{\mathcal{D}}{\rm 3.3\,Gpc}\right]
    \label{eqn:tmulti}
\end{equation}
where $\thE$ is the Einstein radius, $\mup$ is the combined lens magnification encountered by the two images, $\mathcal{D}=\dlC\dsC/\dlsC$, where $\dlC$, $\dsC$, $\dlsC$ are the comoving distances from the observer to the lens, from the observer to the source, and from the lens to the source respectively. Assuming values of lens redshift $\zl=0.5$, which corresponds to the peak of optical depth to strong lensing \citep{robertson19}, and source redshift $\zs=1.6$, which corresponds to the predicted peak of the true distances of the lensed BNS population \citep{smith21}, yields $\mathcal{D}=3.3\,\rm Gpc$. We present an application of these estimates in Section~\ref{sec:time-delay}, when we consider lenses for which the density profile matches that of an SIS at the Einstein radius.

\subsection{Kilonova Lightcurves}\label{sec:kilonova_models}

In parallel with our observing campaign, new models for KN lightcurves were being developed following the extensive follow-up efforts on KN AT2017gfo associated to GW170817. We take advantage of this progress to revisit the sensitivity of our observations. In particular, we use KN models from \citet{nicholl21} to predict the time evolution of the apparent magnitude of the EM counterpart to GW events. These models take as input parameters properties of GW mergers that can be recovered directly from the detected waveform, i.e. chirp mass, binary mass ratio, and inclination angle, as well as the equation-of-state dependent parameters of tidal deformability and maximum NS mass. 

\begin{figure}
\begin{center}
    \includegraphics[width=\linewidth, keepaspectratio]{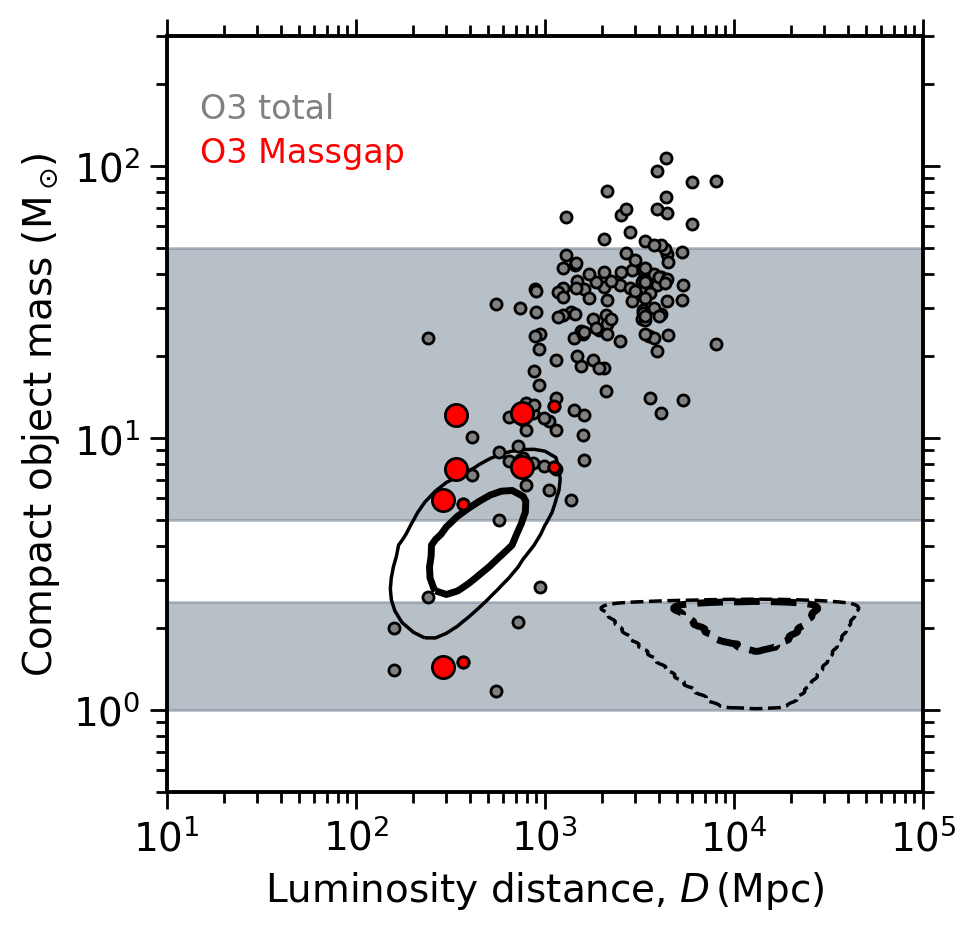}
 \end{center}
  \caption{Distributions of luminosity distance and mass for BNS models, and for detected GW events in O3, following \citet{smith21}. Filled points mark the median posterior mass-distances of the individual binary components that LVK infer assuming $\mu=1$ for the O3 detections. Red points mark the events classified as mass gap with low latency by LVK. The large red points mark the individual component masses of the progenitors of the GW events considered in this study.} Dashed contours show the intrinsic mass-distance distribution expected for the lensed BNS population detected by LIGO-like instruments, while the solid contours show the inferred mass-distance distribution (assuming no lensing) for the same population. 
  In each case the thicker (inner) and thinner (outer) contours encircle $50$ and $90$ per cent of the predicted magnified population respectively. Grey horizontal bands show the mass range for typical NSs and stellar BHs.  \label{fig:contours}
\end{figure}

We give a brief overview of \citeauthor{nicholl21}'s models and refer the reader to their paper for full details. The electromagnetic emission from a KN event is constructed using five main ingredients. The first component describes the mass ejecta due to dynamical forces during the binary merger and is modelled using a highly opaque, equatorial component (labelled ``red'' and responsible for the production of heavy elements). The second bluer, polar ejecta is caused by shocks at the binary contact surface and is characterised by a $\approx$ 20 times lower opacity compared to the red ejecta. The relative contribution of red and blue ejecta is most sensitive to the mass ratio of the binary system. The third component describes the mass ejecta that follows the binary merger, resulting from the interaction of neutrino winds with diffuse merger remnants, which are typically more massive than the pre-merger ejecta, and characterise by an opacity that lies between the red and blue dynamical ejecta. In summary, the geometry of the tri-phase ejecta is approximated by a sphere dominated by the red component for angles $|\theta|<45$deg from the orbital plane, and the lower opacity components carving a bi-conical polar cap orthogonal to the orbital plane. A fourth model component accounts for enhanced emission of blue ejecta due to magnetically-driven winds. The total output luminosity driven by r-process decay is proportional to time $t^{-1.3},$ and is computed considering each ejecta component mass, together with its velocity and opacity, which is assumed constant. An additional fifth component allows for the onset of a gamma-ray burst (GRB) jet which can be responsible for additional shock-heating of the ejecta.

\begin{figure*}
 \begin{center}
    \includegraphics[width=\linewidth]{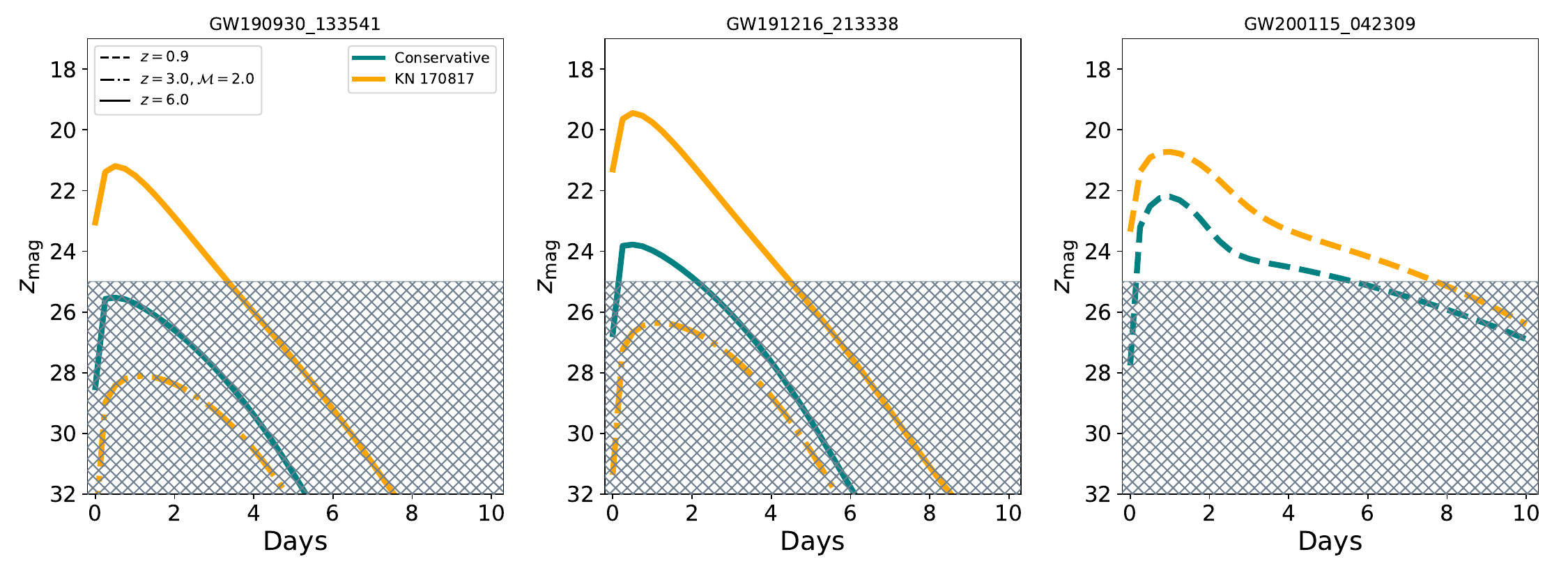}
 \end{center}
  \caption{Predicted evolution of $z'$-band magnitudes for KN emission following the models by \citet{nicholl21} for each of the GW events considered here. The reference model, labelled KN170817, is plotted as yellow lines and uses the best fit parameters to the KN observed in conjunction with GW 170817 \citep{nicholl21}. In addition, we test lightcurves using the reference model parameters and $\mathcal{M}=2 \, M_\odot$, plotted as dot-dashed lines. A conservative model is included for comparison as teal lines. For this, we assume a merger with chirp mass $\mathcal{M}=1.18 \, M_\odot$, mass ratio $q=0.9$, viewing angle $\theta = 60 \,$deg, and no additional blue ejecta from magnetic winds, or shock cooling emission.  The magnitude values are boosted according to the ratio between the LVK posterior distances, which assume no lensing, and a range of redshifts $z_{\rm true}\in[0.9, 3.0, 6.0]$, marked with dashed, solid and dot-dashed line respectively, and assumed as the true location of the binary merger. The gray area marks magnitudes beyond the reach of our observations}. \label{fig:lightcurves}
\end{figure*}

\section{Results}\label{sec:results}
We have continued the deep search for EM counterparts to candidate lensed GW events started by \citet{smith19a}. Our campaign targeting the appearance of new transient phenomena within the central, strong lensing, regions of three galaxy clusters yielded no significant candidates. Thanks to the inclusion of prescriptions for BNS lensing and of state-of-the-art KN lightcurve models we are able to further inform our strategy and findings, and to discuss the implications of lensing on the posterior parameters from LVK.

Figure~\ref{fig:contours} shows the true and lensed distribution of mass and distance for the model BNS population discussed in Section~\ref{sec:gw_model}. We note that  lensed BNS are predicted to be located primarily within the boundaries of the mass gap interval, confirming this as a viable mechanism responsible for the detection of binary components with masses inferred to be between $2.5<m \, [{\rm M_\odot}]<5$. The median of the posteriors of mass and distance for individual binary components from confirmed GW events from LVK's third run are marked as filled circles, and include the events considered in this study in red (see Section~\ref{sec:massgap_events}). We note that the final mass posteriors released from LVK for the mass gap events considered here have values outside of the mass gap. This is due to the different offline pipeline deployed by LVK, with respect to the ones used for the initial detection and classification of the GW trigger. This strengthens the case for an improved analysis of the GW events with low latency from LVK, which would yield robust mass posteriors informing the selection of the most promising events to follow-up. Nevertheless, we note that the predicted lensed population of BNS extend beyond the mass gap, and that the mass-distance posteriors of several progenitors of the events considered here are consistent with the lensing scenario (see overlap between red points and solid contours in Figure~\ref{fig:contours}.

Lensing allows us to reconcile the mass posterior from LVK to that expected for the GW sources considered here by adjusting the true redshift of the merger. We adopt a true chirp mass $\mathcal{M} = (m_1 m_2)^{3/5}(m_1+m_2)^{-1/5} = 1.18\,{M}_{\odot}$, which corresponds to the best-fit describing the KN AT2017gfo observed in conjunction with GW170817 \citep{nicholl21}. Following $\mathcal{M}(1+z)=\tM\,(1+\tz)$, the true redshifts of GW 190930$\_$133541, GW191216$\_$213338 and GW 200115$\_$042309 would be $z = [6.0, 6.0, 0.9]$, respectively. We note that this would yield extreme magnification ($\mu > 3000$) for GW 190930$\_$133541 and GW191216$\_$213338, compared to the predicted values for O3, whose distribution peaks at $\mu \approx 1000$ \citep{smith21}. Therefore, we also consider a higher intrinsic chirp mass $\mathcal{M}[M_{\odot}] = 2$ for these two events, which translates into individual binary masses $m_1=m_2=2.3\,M_{\odot}$ when considering equal mass ratio merger, and a true redshift $z\approx3.0$. This increase in true intrinsic chirp mass is within the range of NS masses inferred by \citetalias{lvc21}.

\subsection{Expected lightcurves}
Models for lightcurves require the input of parameters describing the BNS merger and remnants. As delineated above, we use the best-fit parameters from the AT2017gfo counterpart to GW170817 to inform our choice of parameters. This is so as to produce an empirically-motivated set of light curves. Specifically, we adopt an intrinsic chirp mass $\mathcal{M} = 1.18 \,{\rm M_\odot}$, mass ratio $q = 0.92$, a shocked cocoon opening angle $c =24$ deg, blue ejecta enhanced by a factor 1.6, post-merger ejecta of $12\%$ of the total disc mass, and a viewing angle of $\theta = 32$ deg, to our reference lightcurve model. We produce two additional models: one with the same parameters but a larger chirp mass $\mathcal{M}  = 2.0\,{\rm M_\odot}$; and a "conservative" model with typical NS binary parameters $\mathcal{M}  = 1.18\,{\rm M_\odot}$, $q = 0.9$, viewed at $\theta = 60$ deg, with no shock cooling or blue ejecta enhancement.

Figure~\ref{fig:lightcurves} shows the different model KN lightcurves for each of the three GW events considered in this study. Overall, we note that GW170817-like models predict a detectable transient with our follow-up strategy, together with highlighting the necessity of quick activation given the rapid fading of the transient magnitude within $\simeq4$ days of the merger. Both GW 190930$\_$133541 and GW 191216$\_$213338 present short spans of visibility, due to the high true redshift required to reconcile the posterior chirp mass to values compatible to that of BNS. Hence, the high magnification results in short, high luminosity peaks associated with rest-frame ultraviolet emission that rapidly fades in the observer frame notwithstanding the cosmological time dilation. Lightcurves for GW200115$\_$042309 allow visibility up to 8 days with the current magnitude limits from our campaign, benefitting from relatively low true redshift ensuring that the observations would have probed the rest frame optical emission around $500\,\rm nm$. A common feature of all the lightcurves presented here is that their duration, defined as the time required to fade by a factor 2 in flux, is below 2 days. This makes them the fastest fading transients and suggest short duration as key signature of lensed KNe (for additional details see \citealt{smith21}).

We note that the GW170817-like model is brighter than the conservative model by 2-3 magnitudes, owing to the combined effects of larger blue ejecta mass, favourable viewing angle, and cooling emission from the shocked cocoon. These two models therefore provide optimistic (but plausible) and more pessimistic scenarios for EM counterparts. When considering the implication of varying the true redshift of the merger, we note that the steep decrease of the blue component of the KN, which is responsible for the luminosity peak, is longer lasting in the observer frame than the rest frame due to cosmological time dilation. This effect delays the onset of the lightcurve red component, and accentuates the fading of the observed magnitudes. In this respect, the red component is observable for the closer event considered here GW 200115$\_$042309 when assuming a true redshift $z=0.9$.  Regarding GW 190930$\_$133541 and GW 191216$\_$213338, using $\mathcal{M}_{\rm}= 2\,{\rm M_\odot}$ results in a decrease of the overall luminosity due to the significant reduction of merger ejecta, due to the rapid collapse of the massive merger product to a black hole \citep{nicholl21}.

\subsection{Multiple image interpretation}\label{sec:time-delay}

We now turn to a more detailed discussion of GW190930$\_$133541 and GW191216$\_$213338.
The overlap between the skymaps associated with
these detections, in which cluster RX\,J2129.6$+$0005 is included,  motivated us to discuss the hypothesis that they are the manifestations of a single, gravitationally lensed event. In addition, the median values of merger mass ratio $q$, which is lensing-invariant, are consistent between the two events, but are typically not well constrained \citep{gwtc2}.
Whilst this does not provide conclusive evidence of strong lensing, we utilise the difference between the arrival time of these two GWs ($\approx78$ d) to explore the hypothesis that they are strongly lensed images of the same source.

The ratio between these events' peak posterior distances is $\tD ({\rm GW} 190930)/\tD ({\rm GW} 191216) \approx 2$, which implies a magnification ratio of $\mu ({\rm GW} 191216)/\mu({\rm GW} 190930) \approx 4$. This is firmly in the regime of low-magnification strong lensing, in which isothermal galaxy-scale lenses are more efficient than cluster-scale lenses \citep{smith21}. We therefore adopt the following expression under the assumption that the putative lens is close to isothermal:
\begin{equation}
    \mu_{\pm} = 1\pm \thE\, \beta^{-1},
\end{equation}
where $\beta$ is the source position and the respective images are denoted "$+$" and "$-$". This 
enables us to estimate the individual magnifications suffered by the two images as $\mu_+ \approx 2.67$ and $|\mu_{-}| \approx 0.67$, and thus a combined magnification of $\mu_p \approx 3.3$. This shows that GW 190930$\_$133541 and  191216$\_$213338 are de-magnified and magnified, respectively. The true distance of the source can be rewritten as:
\begin{equation}
    D  = \tD_+ \mu_+^{1/2} = \tD_- |\mu_-|^{1/2},
\end{equation}
which yields $D \approx600\,{\rm Mpc}$ which corresponds to $z\approx0.13$. Given the low redshift, the GW posteriors are only marginally affected. Then substituting $\mu_{\rm p}\simeq3.3$, $\Delta t=78\,\rm d$ and $\mathcal{D}\approx D\approx600\,\rm Mpc$ in to Equation~\ref{eqn:tdouble}, we obtain a $\thE\approx2$ arcsec.  The following relation for an isothermal lens
\begin{equation}
  \thE=\frac{4\pi\sigma^2}{c^2}\frac{\dls}{\ds}
\end{equation}
then allows us to estimate the velocity dispersion of the lens as $\sigma = 370 \,{\rm km \,s^{-1}}$, assuming that $\dls=\ds/2$. This value is typical of small group-sized halos that are commonly inhabited by a massive early type galaxy -- i.e. a fairly typical low magnification strong lens configuration. Repeating this exercise to characterise the properties of a lens producing more than two images using Equation~\ref{eqn:tmulti} yields an Einstein radius is $\thE = 7.8$ arcsec and $\sigma = 735\,{\rm km \,s^{-1}}$, which is typical of a cluster-scale lens, and thus less plausible given the low magnifications at play here \citep{smith21}.

\section{Conclusions}\label{sec:conclusions}
The steady progress in the detection of GWs revealed growing inconsistency in the understanding of the mass regimes accessible to compact objects. The detection of mass gap objects during LVK's third run highlights this tension. The intervention of gravitational lensing offers a physically well-understood solution to the detection of these GW events, without invoking modifications to the current models of stellar evolution. To test this hypothesis, we have selected promising mass gap events detected during LVK's third run, and performed a search for EM counterparts to explore if they were affected by strong lensing. In particular, we have selected three GW events as containing at least one component whose mass falls within the mass gap $2.5<m \,[\rm{M_{\odot}}]<5$, and within hours of their detection, selected and observed one galaxy cluster within each event localisation map using Gemini-GMOS. The data obtained reach magnitude $m_z \approx 25.5$ at $5\sigma$, and we detect no clear transient within the data limit, when comparing the newly obtained observations to deep \emph{HST} archival imaging. This is the continuation of the deep pilot search for EM transients to GW events started by \citet{smith19a} in anticipation of the upcoming opportunities afforded by the Vera Rubin Observatory.

Thanks to state-of-the-art lightcurve models for KN-like transients, we are able to confirm that EM counterparts to lensed GW events would have been detectable thanks to lensing magnification using 8-meter class ground-based telescope within days from the binary mergers. Lensing arguments highlight that posterior values of chirp masses up to $\tM\approx9 M_{\sun}$ can be compatible with that of binary neutron star mergers. Furthermore, we tested the scenario in which GW 190930$\_$133541 and 191216$\_$213338 are multiple images of a single, multiply imaged gravitationally lensed event. We compared the arrival time difference between the detection of these two events with theoretical predictions of arrival time differences and found that lensing arguments allow for both events to be interpreted as strongly lensed images of an individual GW source multiply imaged by a modest group-scale lens. Given the low magnifications involved, the true masses of the compact objects involved in the merger would be consistent with those inferred by LVK under their assumption that lensing was not at play. 

Looking to the future, significant progress in the search for lensed BNS will arrive with the increase in GW detector sensitivity prospected for LIGO's fifth run, with predicted lensed rates around one per year \citep{smith21}. Our candidate lensed BNS strategy is effective in selecting the most promising GW events affected by lensing, and offers physically well-motivated and sensitive observations of any false positives. Additional progress can be achieved by having access to robust event posteriors on mass and mass ratios with low latency, i.e. as soon as possible after the detection of a GW event, and by increasing the census of lines-of-sight with high lensing probability \citep[][]{ryczanowski22}. This is particularly important in light of LVK's upcoming fourth run (early 2023), which features increased sensitivity, and for which the Vera Rubin Observatory will not be accessible yet. This is essential to select more efficiently the most promising candidates, inform the search for lensing intervention, and focus observational follow-up efforts.

\section*{Acknowledgements}
MB and GPS acknowledge support from STFC grant numbers ST/N021702/1 and ST/S006141/1. DR and ER acknowledge STFC studentships. GPS acknowledges support from The Royal Society and the Leverhulme Trust. MN is supported by the European Research Council (ERC) under the European Union’s Horizon 2020 research and innovation programme (grant agreement No.~948381) and by a Fellowship from the Alan Turing Institute.
We thank the staff and Directors of the Gemini and UKIRT Observatories and the Cambridge Astronomy Survey Unit for their superb support of our observing programmes. UKIRT is owned by the University of Hawaii (UH) and operated by the UH Institute for Astronomy. When (some of) the data reported here were obtained, the operations were enabled through the cooperation of the East Asian Observatory. Based on observations obtained at the international Gemini Observatory, a program of NSF’s NOIRLab, which is managed by the Association of Universities for Research in Astronomy (AURA) under a cooperative agreement with the National Science Foundation. on behalf of the Gemini Observatory partnership: the National Science Foundation (United States), National Research Council (Canada), Agencia Nacional de Investigaci\'{o}n y Desarrollo (Chile), Ministerio de Ciencia, Tecnolog\'{i}a e Innovaci\'{o}n (Argentina), Minist\'{e}rio da Ci\^{e}ncia, Tecnologia, Inova\c{c}\~{o}es e Comunica\c{c}\~{o}es (Brazil), and Korea Astronomy and Space Science Institute (Republic of Korea). 

\section*{Data Availability}
The Gemini/GMOS dataset used in this work is available at the \href{https://archive.gemini.edu/searchform}{Gemini Observatory Archive}. The UKIRT/WFCAM dataset is available upon request to the authors.



\bibliographystyle{mnras}
\bibliography{biblio.bib} 








\bsp	
\label{lastpage}
\end{document}